\documentclass[aps,pra,reprint,twocolumn,showpacs,floatfix,superscriptaddress]{revtex4-1}

\usepackage{amssymb,amsmath,amstext}                
\usepackage{graphicx}                                               
\usepackage{epstopdf}                                               
\usepackage{color}                                                     
\usepackage{bm}                                                        
\usepackage{appendix}                                              
\usepackage[utf8]{inputenc}
\usepackage{bbold}
\usepackage{bbm}
\usepackage{latexsym}
\usepackage{xcolor}
\usepackage{braket}
\usepackage{ulem}
\usepackage{wasysym}
\usepackage{verbatim}

\definecolor{lblue} {RGB}{51,71,158}
\usepackage[colorlinks=true,citecolor=blue,linkcolor=blue,urlcolor=lblue]{hyperref}

\newcommand{\average}[1]{\overline{#1}}

\begin{document}

\title{{Many-body localization of bosons in optical lattice: \\
Dynamics in disorder-free potentials}}
\author{Ruixiao Yao}
\affiliation{School of Physics, Peking University, Beijing 100871, China}
\author{Jakub Zakrzewski}
\email{jakub.zakrzewski@uj.edu.pl}
\affiliation{Institute of Theoretical Physics, Jagiellonian University in Krakow,  \L{}ojasiewicza 11, 30-348 Krak\'ow, Poland }
\affiliation{Mark Kac Complex
Systems Research Center, Jagiellonian University in Krakow, Krak\'ow,
Poland. }

\date{\today}

                              
\begin{abstract}
The phenomenon of Many-Body Stark Localization of bosons in  tilted optical lattice is studied. Despite the fact that no disorder is necessary for Stark localization to occur, it is very similar to well known many body localization (MBL) in sufficiently strong disorder. Not only the mean gap ratio reaches poissonian value as characteristic for localized situations but also the eigenstates reveal multifractal character as in standard MBL. Stark localization enables a coexistence of spacially separated thermal and localized phases in the harmonic trap similarly to fermions. Stark localization may also lead to spectacular trapping of   particles in  a reversed harmonic field which naively 
might be  considered as an unstable configuration.
\end{abstract}

\maketitle
\section{Introduction}
Nature seems not to like boredom. When Fermi, Pasta and Ulam wanted to test an approach to ergodicity in nonlinear systems, they found with M. Tsingou, in a famous numerical experiment on coupled nonlinear oscillators, a quasiperiodic motion with energy shared among few modes only (see e.g. \cite{Ford92}). The strongly chaotic classically motion 
leads to spectra with statistical properties well described by random matrix theory \cite{Haake} as revealed e.g. by hydrogen atom in a uniform magnetic field. Yet, a closer both experimental and theoretical analysis
revealed system specific regularities traced back to classical periodic orbits (for a review see \cite{Friedrich89}) and semiclassical quantization schemes based on them \cite{Gutzwiller71}
received a deserved verification. Similarly it was believed for a long time that  generic  many-body systems in their time evolution are faithful to  Eigenstate Thermalization Hypothesis (ETH) \cite{Deutsch91,Srednicki94,Alessio16} with notable exceptions of integrable systems which can be described by generalized Gibbs ensemble -- see e.g. \cite{Cassidy11,Rigol12}. In one sentence, {under ETH} {few-body} observables are expected to thermalize due to interactions within the whole system. A powerful counterexample was found for strongly disordered systems \cite{Gornyi05,Basko06} which was later coined 
many-body localization (MBL). MBL, a robust ergodicity breaking phenomenon, has been intensively studies since then (for reviews see \cite{Huse14, Nandkishore15, Alet18, Abanin19}). Theoretical studies mainly concentrated on interacting spin systems but experimental demonstration of MBL came with  cold atom platforms
\cite{Schreiber15,Luschen17,Lukin19,Rispoli19}. While experiments deal with finite systems, very recently questions about the very existence of MBL in the thermodynamic limit
have been posed \cite{Suntajs19} initiating a vivid discussion  \cite{ Abanin19a, Sierant20b, Panda19,Suntajs20,Laflorencie20,Sierant20c}. Significant progress has been made in addressing dynamics of large systems
\cite{Zakrzewski18,goto_prb_2019, Doggen19,Chanda20t,Chanda20m}. In here we shall limit ourselves mostly, however,  to small systems in the spirit of recent experiments with bosons \cite{Lukin19,Rispoli19} for which simulations may be performed numerically exactly \cite{Yao20}.  {Only for slowly varying harmonic potential we consider larger system sizes using matrix product states formalism \cite{Schollwoeck11}.}
\par
MBL is not the only ergodicity breaking mechanism identified recently. Interacting Rydberg atom arrays revealed persistent oscillations  \cite{Turner18,Wen19, Khemani19, Iadecola19, Iadecola19a} sometimes referred to as quantum scars. However, in traditional quantum chaos language quantum scars denote partial wavefunction localization on unstable periodic orbits \cite{Heller84,Bogomolny88} - as revealed by the mentioned above hydrogen atom in magnetic field problem. The oscillations observed for Rydberg atoms does not seem to have any classical counterpart (see e.g. \cite{Mark20,Michailidis20,Turner20}). Similarly nonergodic behavior has been observed for models with global constrains, notably lattice gauge theory models  \cite{Smith17, Brenes18, Magnifico19, James19,
Chanda20, Giudici19, Surace19,Feldmeier20} or for fragmented Hilbert space  \cite{Sala20,Khemani19b, Rakovszky20}. Recently, strong nonergodic behavior has been predicted for a static uniform force acting on atoms in an optical lattice \cite{Schultz19,vanNieuwenburg19,Taylor19}.  The  phenomenon has been naturally called  a many-body  Stark Localization and has been recently discussed in spinless and spinful fermions
 \cite{Schultz19, vanNieuwenburg19, Chanda20b}. The same mechanism has been found to be responsible for the predicted spacial coexistence of extended and localized regions \cite{Chanda20b}. While many-body Stark localization has been considered for spins and fermions, for bosons it was mentioned briefly only \cite{Taylor19}, the aim of this paper is to fill this gap.
 \par
 MBL of bosons in disordered potential was addressed in several works \cite{Aleiner09,Aleiner10,Michal16}. In optical lattice it shows characteristic features related to the fact that there is no limitation for number of particles occupying a single site. Thus one may expect additional effects due to bunching as, e.g.,  the existence of the inverse mobility edge, with higher lying states being easier to localized for sufficiently strong interactions \cite{Sierant17b,Sierant18,Hopjan19,Yao20}. Also random interactions \cite{Sierant17} as well as cavity-mediated long-range interactions \cite{Sierant19c} were considered in the context of bosonic MBL.
In this paper, we address quantitatively the many-body Stark localization for bosons in the absence of disorder assuming a tilted optical lattice within the  standard Bose-Hubbard model.
In Section~\ref{spectral} we demonstrate the existence of localized phase and determine the critical field strengths for the crossover at finite system sizes considered. We analyse also the properties of eigenstates finding them to be multifractal. This property is shared with standard MBL \cite{Mace19} as well as with ground state features \cite{Lindinger19} of the model.
Later, we consider effects due to an additional  harmonic, slowly varying potential  on  top of the optical lattice that allows us to study the coexistence of both thermal and localized phases in similarity to fermions \cite{Chanda20b}.Finally we show that the fact that the curvature of external potential could suppress transport may lead to the possibility that atoms might be trapped by a reversed harmonic field.
\par

\section{The Hamiltonian and its spectral properties}
\label{spectral}

We shall {consider} bosons confined in a quasi one-dimensional optical lattice. We describe the model by a standard Bose-Hubbard Hamiltonian:
\begin{equation} 
\hat H=-J\sum_{k}^{L-1}(\hat{b}_k^{\dagger}\hat{b}_{k+1} + h.c.) + \frac{U}{2}\sum_{k}^{L}\hat{n}_k(\hat{n}_k-1) + \sum_{k}\mu_k\hat{n}_k
\label{boson}  
\end{equation}
where $\hat{b}_k(\hat{b}_k^{\dagger})$ denote bosonic annihilator (creator)  operators obeying commutation relation $[\hat{b}_k, \hat{b}_t^{\dagger}] = \delta_{kt}$ and $\hat{n}_k = \hat{b}_k^{\dagger}\hat{b}_k$. For standard MBL studies \cite{Sierant18} one assumes $\mu_k$ to be random. Instead we shall consider the system  {in a} tilted lattice with  the on-site chemical potential of the form of $\mu_k = Fk$ which corresponds to a uniform force acting on
 bosons. From now on we assume $J=1$ (with $J$ being thus a unit of energy).
 We consider small system sizes with number of sites of the order of $L=10$ inspired by recent experiments \cite{Lukin19,Rispoli19}.
\par

It has been found recently \cite{Schultz19,vanNieuwenburg19} that a uniform tilt of the lattice (corresponding to an action of the static electric field) may lead, for spinless fermions, to disorder free
localization called many-body Stark localization (MBSL) for a sufficiently large electric field, $F$. Soon the corresponding effect has been addressed with experimentally much easier to realize spinful fermions in the tilted lattice case \cite{Bloch20,Chanda20b}. We shall consider it here for bosons where 
no additional restrictions due  to Pauli exclusion principle
are present. As for fermions, MBSL may be viewed as a generalization of Wannier-Stark localization \cite{Emin87,Gluck02} to the interacting case.

\begin{figure}
\includegraphics[width=1.0\linewidth]{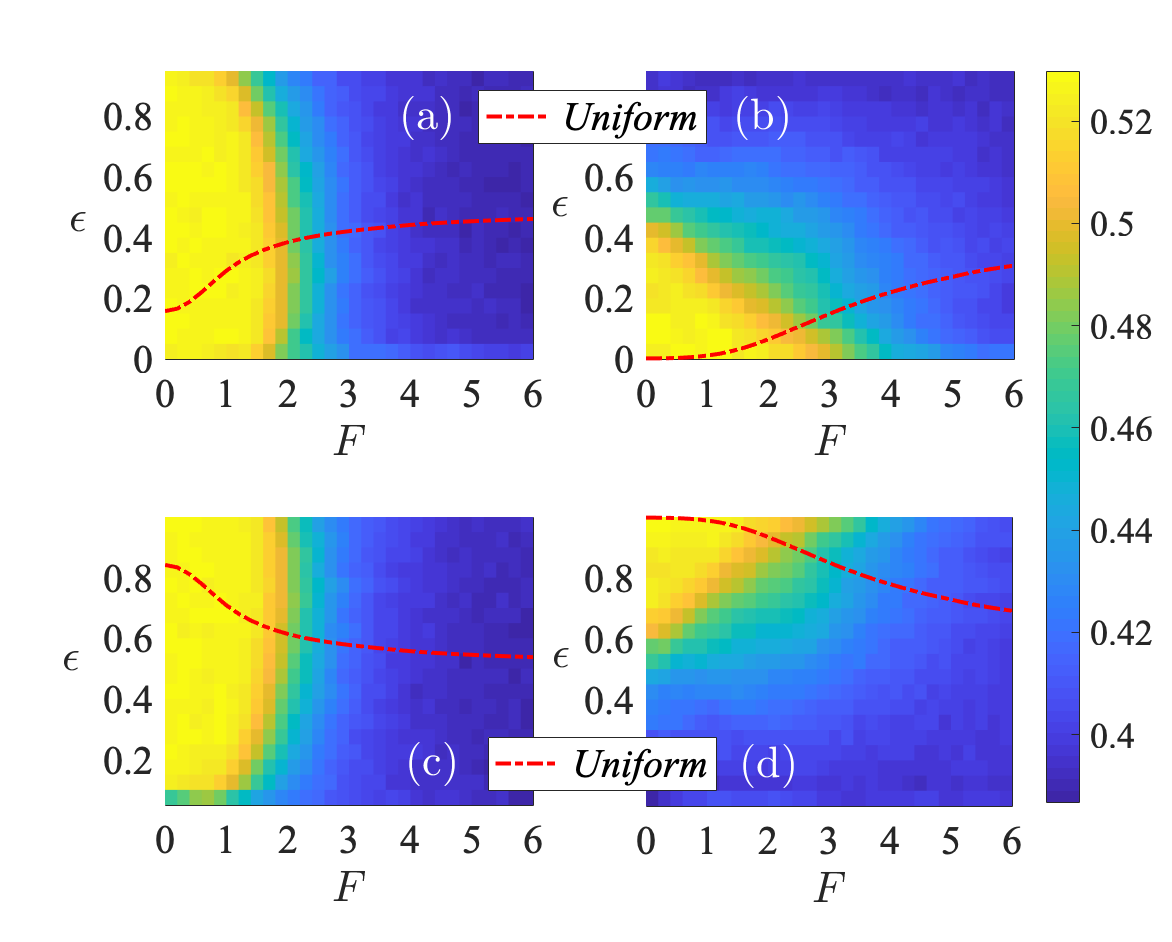}
\caption{{Mean gap ratio $\average r$ for the tilted Bose-Hubbard chain of  $M=8$ bosons on $L=8$ sites in the plane  the scaled energy $\epsilon$ versus the tilt amplitude  $F$.
The red curve  indicates scaled energies of initial states with uniform occupation of sites. The crossover from extended to localized regime depends on energy, quite strongly for $U=5$. 
For attractive interactions  (bottom panels) the energy scale is simply reversed as compared to $U>0$ case (top) and in particular for $U=-5$ ($U=5$) low (high) lying states are localized for arbitrary $F$. This energy range corresponds to the low density region of states with single sites being multiply occupied. 
    \label{rbarheis}
}}
\end{figure}

We shall consider mainly the repulsive interactions, $U>0$. One might expect that for attractive interactions ($U<0$) bosons will tent to group together. This intuition is valid for very low lying energy states, however, we are interested in the properties of highly excited states in the regime of high density of states. Recall that, for $F=0$, the change of the sign of $U$
corresponds to an effective change of the sign of the Hamiltonian (and all the eigenvalues) $H\rightarrow -H$) as the sign in front of $J$ may be simultaneously changed by the gauge transformation $b_k \rightarrow (-1)^kb_k$. The same property holds in the presence of $F$. Due to the total particle number conservation $\sum n_k=const.$ the chemical potential term $Fk$ may be made symmetric around the center of the chain and {the} change of $U \rightarrow -U$ may be accompanied with {the} chain reflection around the chain centre. Thus spectra
of $H$ for  given $J,F$ values are the same for $U$ and $-U$. {This is an exact symmetry between two Hamiltonians with changed $U$ sign and reversed energy ordering of eigenvalues and eigenvectors}.

With this said, let us consider first the spectral properties: eigenvalues and eigenvectors of the model \eqref{boson} with tilt. While in Ref.\cite{Schultz19} a small harmonic potential is added to a pure linear tilt, we shall follow \cite{vanNieuwenburg19} and modify the chemical potential adding a small disorder. In effect, we assume in this section
\begin{equation}
\mu_i=Fi+W(h_i-0.5)
\label{chempot}
\end{equation}
where $h_i$ are random, drawn from a uniform distribution on  $[0,1]$ interval. We take $W=0.5$ which is sufficiently small that in the absence of the tilt, $F$, the statistics is faithful to
Gaussian orthogonal ensemble (GOE) of random matrices \cite{Yao20} for parameters considered. We assume some non-zero disorder to be able to increase statistically the sample of eigenenergies we consider. Typically we consider 200 disorder realizations.

 As a simple indicator of the transition between thermal and localized phases, the mean gap ratio $\bar{r}$ is often used \cite{Oganesyan07,Luitz15,Mondaini15,Sierant19}. It is an average of dimensionless gap ratios defined as 
\begin{equation} 
r_n = min\{\frac{s_{n+1}}{s_n}, \frac{s_n}{s_{n+1}}\}
\label{rbar}  
\end{equation}
with $s_n = E_{n+1} - E_{n}$ being the level spacing. Importantly $r_n$ (and thus $\bar{r}$) is dimensionless and the tricky procedure of level unfolding \cite{Gomez02} is not necessary.
On the ergodic side $\bar{r} \approx 0.53$ as appropriate for  Gaussian orthogonal ensemble (GOE) of random matrices while $\bar{r} \approx 0.38$ for Poisson
statistics (PS) describing the localized case \cite{Atas13}. 
Fig.~\ref{rbarheis} shows the mean gap ratio as a function of the lattice tilt (electric field) $F$ for different scaled energies of the system for $M=8$ bosons on $L=8$ sites. Following the seminal treatment of MBL in Heisenberg chain \cite{Luitz15} we define the dimensionless energy $\epsilon=(E-E_{min})/(E_{max}-E_{min})$ to characterize the system. Disorder averaging allows us to
get a decent statistics for 20 bins along $\epsilon$ axis. Left top panel corresponds to small repulsive interaction $U=1$. Observe that for small $F$ the system is delocalized,  
with $\bar{r}$ close to GOE value (except at the very top energy). For larger $F$, $\bar{r}$ decreases in an energy dependent manner where the energies close to the middle of the spectrum are most resistant to effect of $F$ (in the region of largest density of states). Gradually the crossover to a fully localized phase is accomplished around {$F\approx3.5$}. 
Note that the critical region of the transition from GOE-like to Poisson-like $\bar{r}$ values is quite broad as might be expected for a small system size considered.

The picture is markedly different for $U=5$ where states for scaled energies $\epsilon>0.5$ show localized character for all $F$ values. This is related to the splitting of the Hilbert space into subbands for large interactions \cite{Carleo12,Sierant18}  - this large in scaled energy region corresponds to a small number of states (very low density). Interesting physics concentrates for low $\epsilon$ where a strong energy dependence of the localization transition occurs (the situation resembles the mobility edge picture studied for a random disorder in \cite{Sierant18,Yao20}). The transition to a fully localized phase is completed  {about} $F\approx 5$.

The bottom panels shows the data for the attractive interactions cases, demonstrating that the spectrum is simply reversed with respect to  $U>0$ situation, as expected on the symmetry consideration grounds mentioned above. For that reason we discuss
the properties of eigenstates for repulsive interactions only.

\begin{figure}
\includegraphics[width=1.0\linewidth]{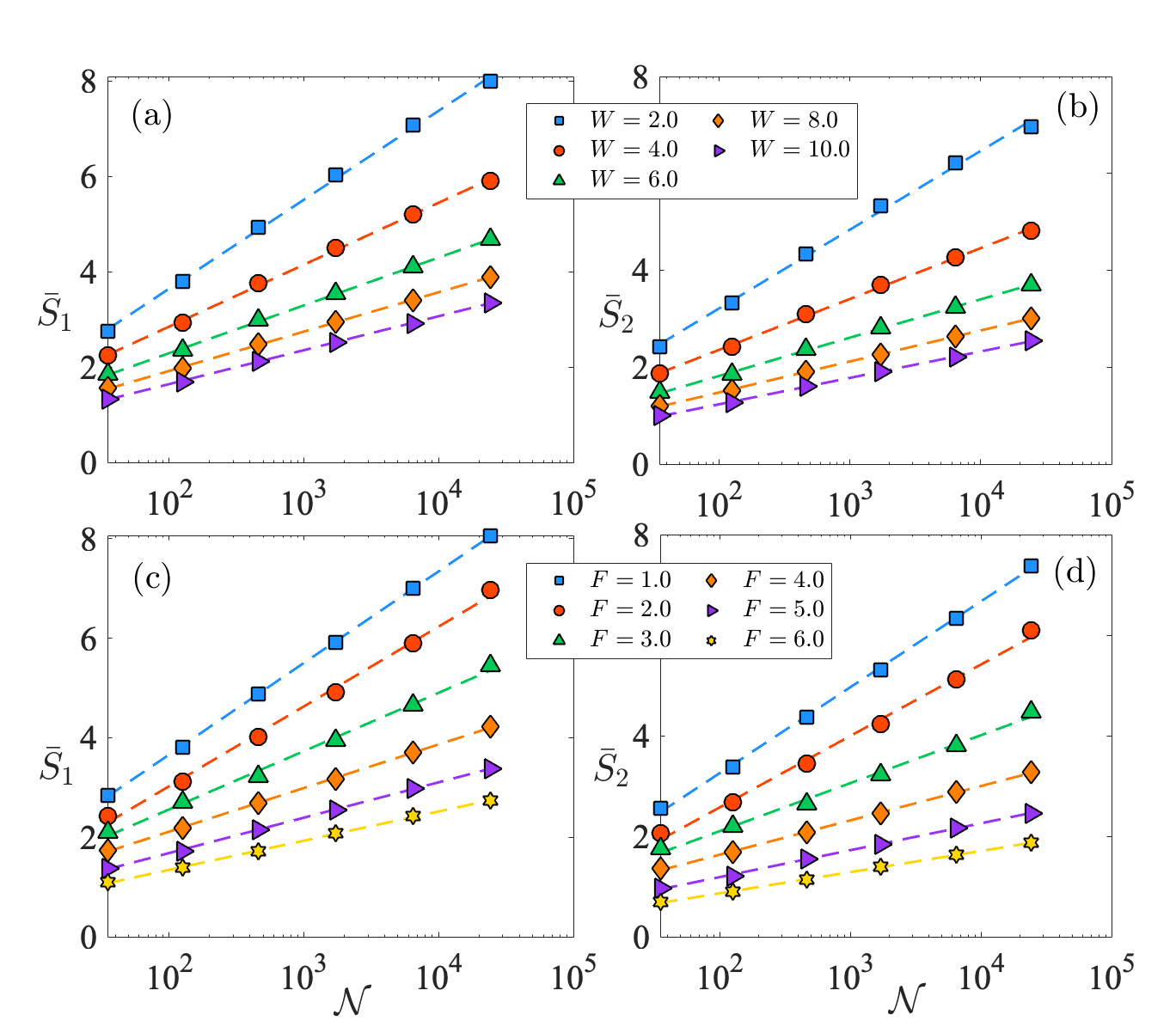}
\caption{{Top: Participation entropies scaling with the system size  for Bose-Hubbard model with the on-site random uniform disorder of amplitude $W$: (a)/(b) correspond to $\bar{S_1}$/$\bar{S_2}$, respectively. The  logarithmic scaling with the Hilbert space dimension ${\mathcal N}$ is apparent. Bottom: the behavior of PEs for the tilted Stark model is very similar with (c)/(d) corresponding to $\bar{S_1}$/$\bar{S_2}$. Data are presented for $U=1$.
    \label{Sqfit} 
}}
\end{figure}

\par The dependence of mean gap ratio on energy and control parameter ($F$) closely resembles the behavior observed for the crossover to MBL for bosons \cite{Yao20} and
spins \cite{Luitz15} alike. One can pose a question - is it just a similarity of statistical properties of eigenvalues or the eigenstates also share similar properties ? For the 
disordered Heisenberg chain a detailed study \cite{Mace18} revealed  multifractality of eigenstates across the critical region as well as in the ``localized'' regime where $\bar{r}$ takes the poissonian value. Let us inspect the properties of eigenstates in the bosonic system. We consider the random on-site disorder case at $F=0$ changing the disorder amplitude $W$ in \eqref{chempot}) (such an analysis for bosons is not available till now) and compare it with Stark-localization  when changing $F$ (with small background disorder $W=0.5$). As in \cite{Mace18} we consider participation entropies (PE) of eigenstates, $S_q$ of order q defined as:
\begin{equation} 
S_q = \frac{1}{1-q}\ln(\sum_{\alpha = 1}^{\mathcal{N}}|\psi_{\alpha}|^{2q}) \ {\mathrm{for}}\ |\Psi\rangle = \sum_{\alpha = 1}^{\mathcal{N}}\psi_{\alpha}|\alpha\rangle
\label{Sq}  
\end{equation}
being an analysed wavefuction. $S_q$ depends of course on the basis $\{|\alpha\rangle\}$ chosen, we restrict to the natural Fock basis below.
We consider changes of $S_q$  with the system size considering  $L = 4$ to $9$. {
In a perfect, GOE like scenario $S_q=\ln {\mathcal N}$ where ${\mathcal N}$ is the dimension of the Hilbert space. on the other hand, for perfectly localized situation $S_q$ should be size independent. In the transition regime one expects PEs to behave like $S_q=D_q\ln\mathcal{N}$
with the fractal dimension $D_q<1$. We call eigenstates multifractal if $D_q$'s are different for different $q$ \cite{Mace18}.

 Fig.~\ref{Sqfit} presents the results of fitting  $\average S_q = D_q\ln(\mathcal{N}) + b_q$ dependence to the data obtained in both studied cases. Data are averaged, as indicated by the overbar, over {1000 disorder realizations except for $L=9$ case with 192 realizations. Typically we take up to 200 eigenstates corresponding to energies around $\epsilon=0.5$ except for smaller $L$ values where necessarily this number is much smaller. For $L=4$ when ${\cal N}=35$ we take just five eigenvalues per realization.}  Still the error bars {obtained} for the {average entropies}  are of the order of symbols sizes. While the system sizes studied are much smaller than for the spin-1/2 Heisenberg case, we observe a striking similarity to the results presented in \cite{Mace18}. It is probably less surprizing that bosons in random disorder reveal a similar trend as spins - after all it is a common believe that general features of MBL phase are similar for different systems. {However, a similarity between the disorder driven (top row) and the electric field driven (bottom row) behavior of $S_q$  apparent from Fig.~\ref{Sqfit}
 is really striking. }
 
{The fitted parameters for $L=8$ sites are collected in Fig.~\ref{Multi}. We observe that $D_1$ and $D_2$ are different from each other for both disorder driven and electric field driven models -- the eigenstates are multifractal even in the regime considered as localized. }
 Moreover, we observe that the $b_q$ free term changes the sign from negative to positive values with increasing disorder (top panel) - a feature also observed for the Heisenberg spin chain \cite{Mace18}. The change of sign of $b_q$ has been identified in \cite{Mace18} as an indicator of the critical disorder value for the transition. In our case $b_q$ changes the sign
 around $W=2$ -- compare Fig.~\ref{Multi} -- while the transition to MBL occurs, according to $\bar{r}$-statistics, for $W>3.5$ \cite{Yao20}  around $\epsilon=0.5$ i.e. the value around which we collect $S_q$ for analysis.  
While this discrepancy is quite significant, let us note large error bars on $b_q$ coefficients
 resulting from the fits, so this discrepancy is within the bounds given by those error bars. Interestingly, the error bars on dimensionality $D_q$ are much smaller and practically within the size of markers in Fig.~\ref{Multi}.

\begin{figure}
\includegraphics[width=1.0\linewidth]{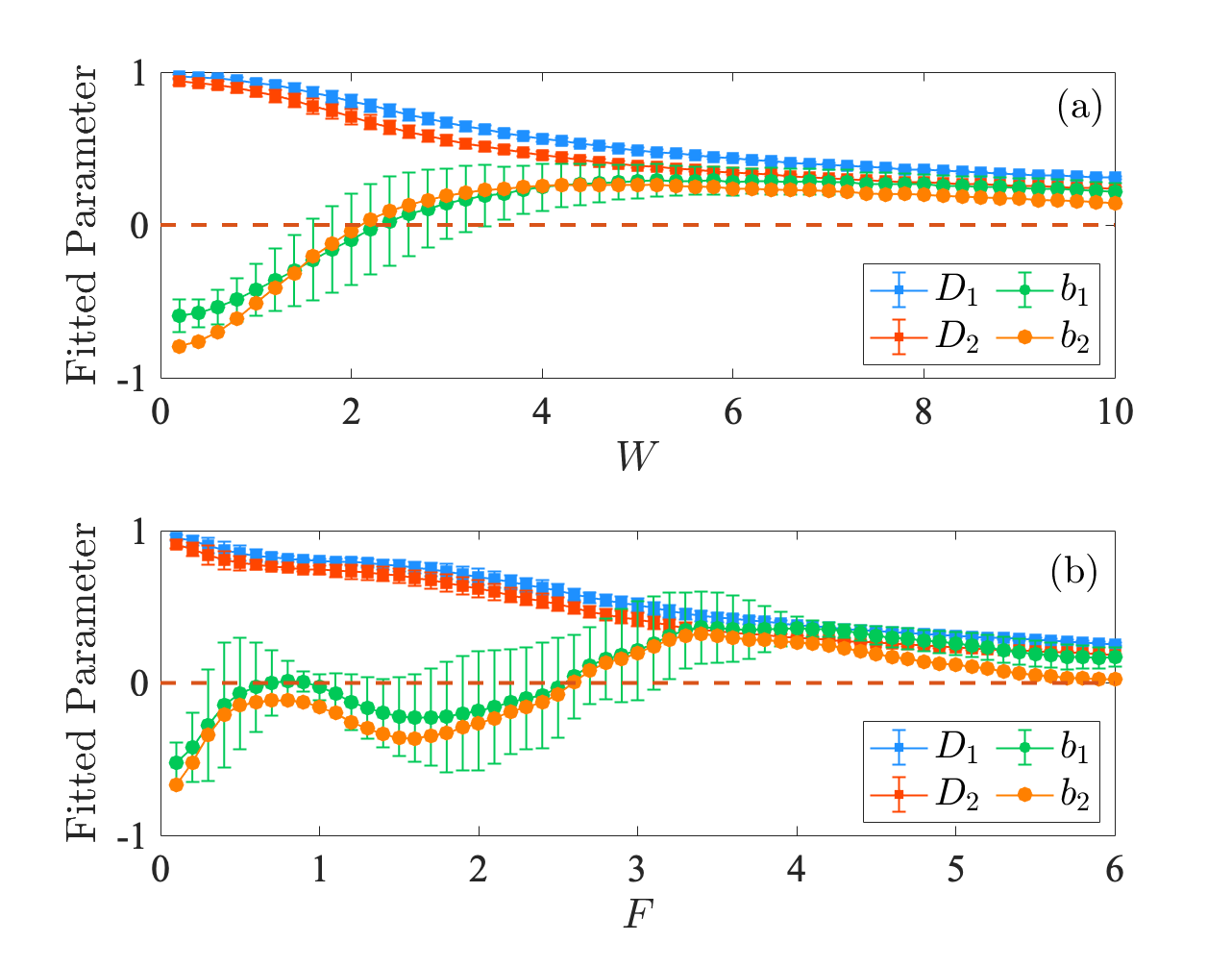}
\caption{{Comparison of fitted  multifractality parameters for standard MBL in bosons as a function of the disorder amplitude $W$ - panel (a) and for many body Stark localization as a function of $F$ - panel (b). Participation entropies, Eq.~\ref{Sq} are fitted as $S_q=D_q\ln{\mathcal N}+b_q$ {for sites $L=4-9$}. The small peak appearing at $F\sim 1$ is possibly the consequence of transport induced by tilt.{The errors for $D_q$ are within the symbol sizes; we show error bars for $b_1$ only for clarity, those for $b_2$ are very similar.}
    \label{Multi} 
}}
\end{figure}

 The bottom row in Fig.~\ref{Multi} analyses the $S_q$ dependence for the tilted lattice. As for spacings analysis, we add a tiny disorder {with amplitude} $W=0.5$ to enable averaging over the disorder realizations. 
 {Similarly to the} random disorder  case, we observe a gradual change of slopes of the $S_q$ fits with increasing tilt, $F$, of the lattice. Even for the largest values of $F$ the multifractal properties of eigenstates persists. Overall the similarity of these two cases of disordered and tilted lattice  suggests that MBL and MBSL are very closely related.
 An interesting behaviour is revealed in the $b_q$ fitted coefficients dependence on $F$ which is not monotonic, in contrast to the random case. 
One observes a fast growth of $b_q$ for small $F$ with the maximum around $F=1$ and a subsequent decrease. Only for much larger $F$ a ``standard''' change of the sign occurs. This behaviour may be related to the fact that the hopping between neighboring sites
  becomes quasi-resonant  around $F=1$ as the additional energy may be supplied by the interaction energy $U=1$ in the case studied. These values of $F$ lead to an enhanced transport in the time dynamics as we shall show below. One should, however, keep in mind a relative large error of $b_q$ coefficients in this regime.

 \begin{figure}
\includegraphics[width=1.0\linewidth]{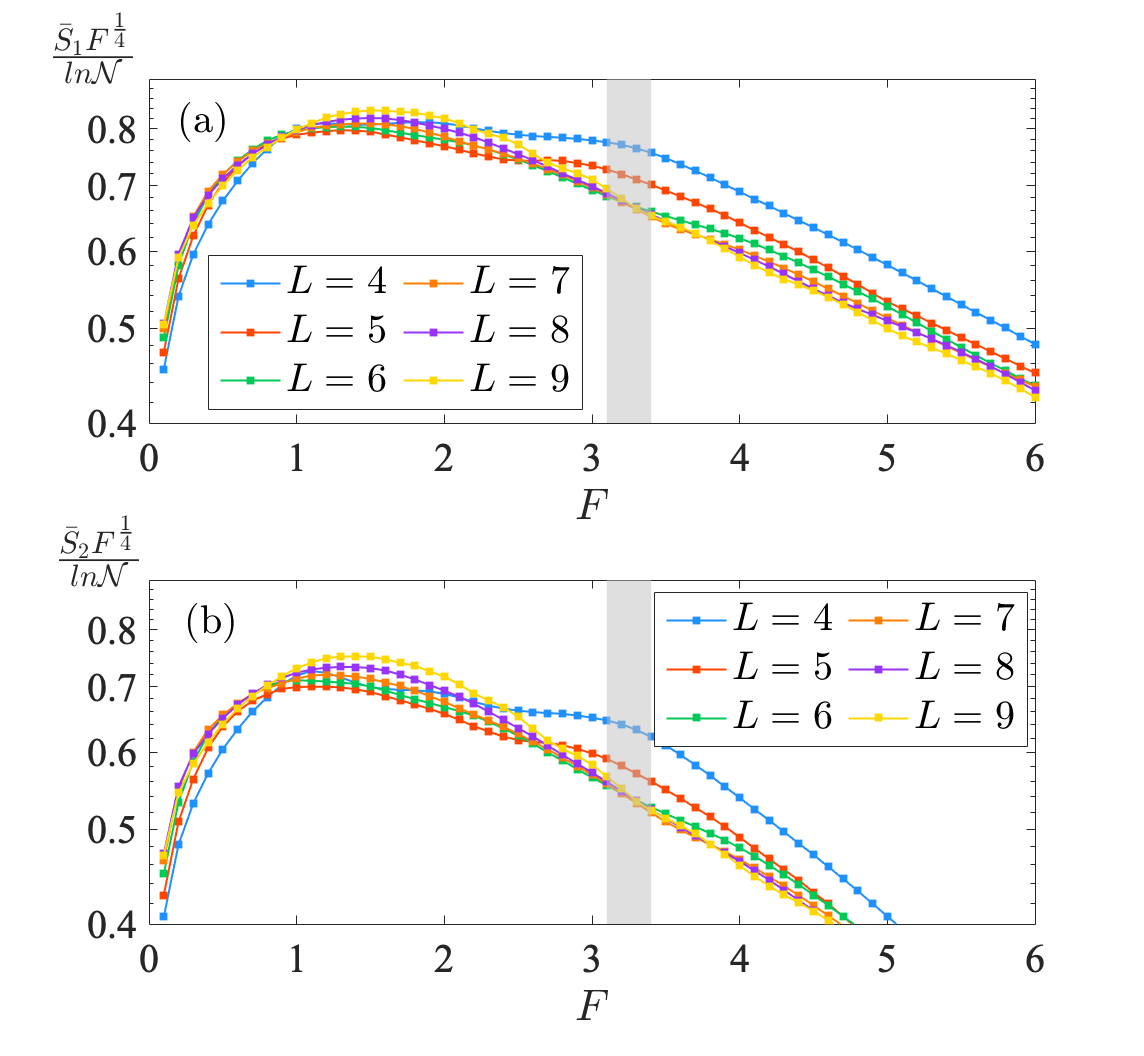}
\caption{{(a)(b): {Rescaled participation entropies, versus field strength. The shaded region indicates the transition occuring at $3.1<F<3.4$ identified by crossing of curves corresponding to different system sizes.  Note that we have  neglected results from small sizes $L = 4,5$  which apparently are affected by small rank of the corresponding matrices.}
    \label{Rescale}
}}
\end{figure}
\par
Rescaling the participation entropies by the logarithm of the Hilbert space dimension (again we precisely follow the scaling analysis of the participation entropy advanced in \cite{Mace18} for the case of spin models)
we may observe a crossing of curves representing $S_q$ for different system sizes - see Fig.~\ref{Rescale}. Multiplication of the data by $F^{1/4}$ factor does not affect the size-dependent crossings but helps  enhancing the details of the crossing.  Both the $S_1$ and the $S_2$ data cross for $F\approx 3.3\pm0.2$ except the data for the smallest system sizes that we disregard. In this way the critical field value, $F_c$, corresponding to the onset of localized regime may be identified.

\section{Transport and Localization}

 \begin{figure}
\includegraphics[width=0.75\linewidth]{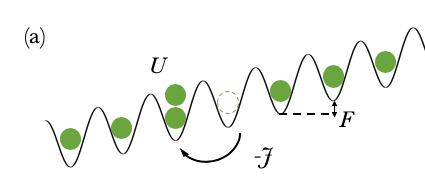}\vspace{-0.2cm}
\includegraphics[width=1\linewidth]{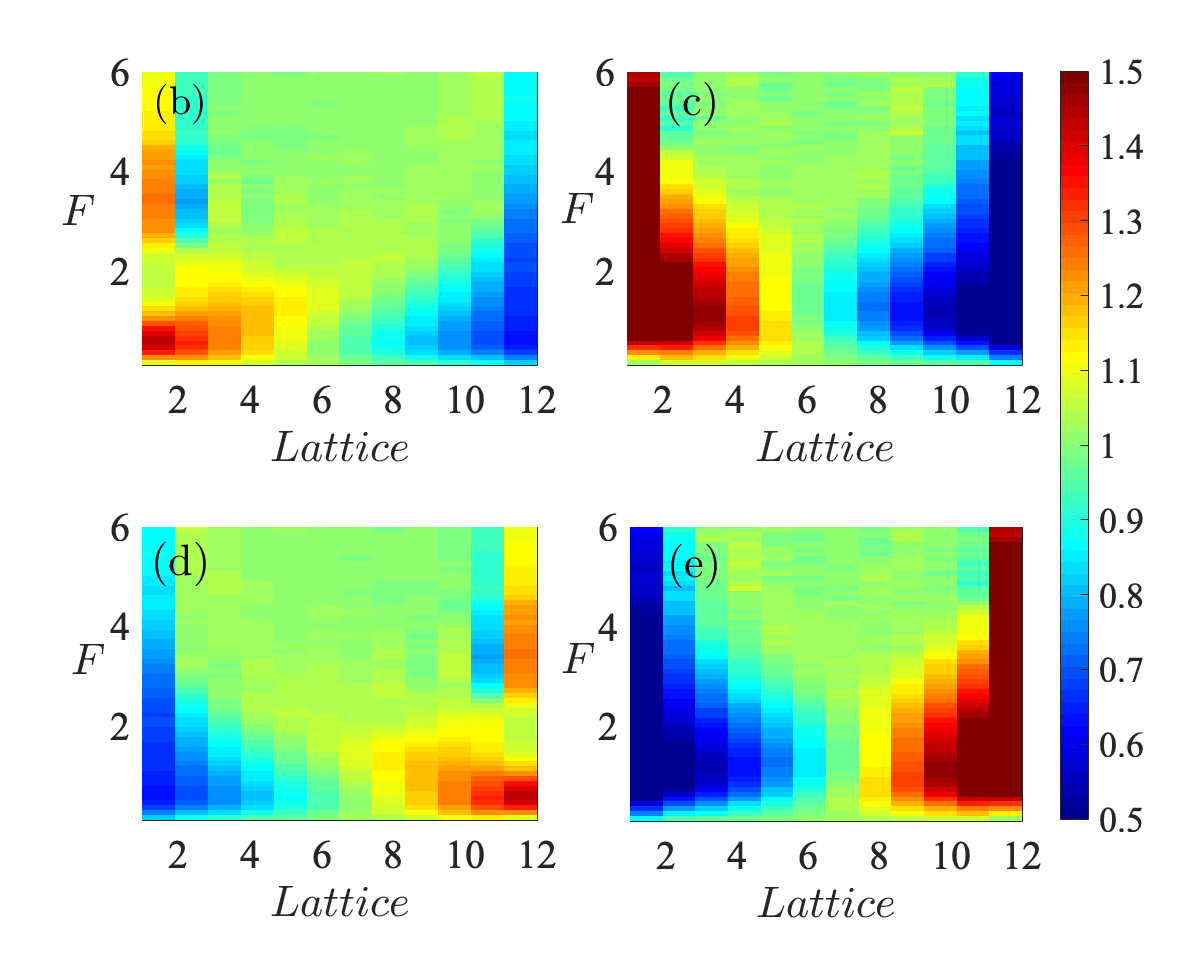}\vspace{-0.2cm}
\includegraphics[width=1\linewidth]{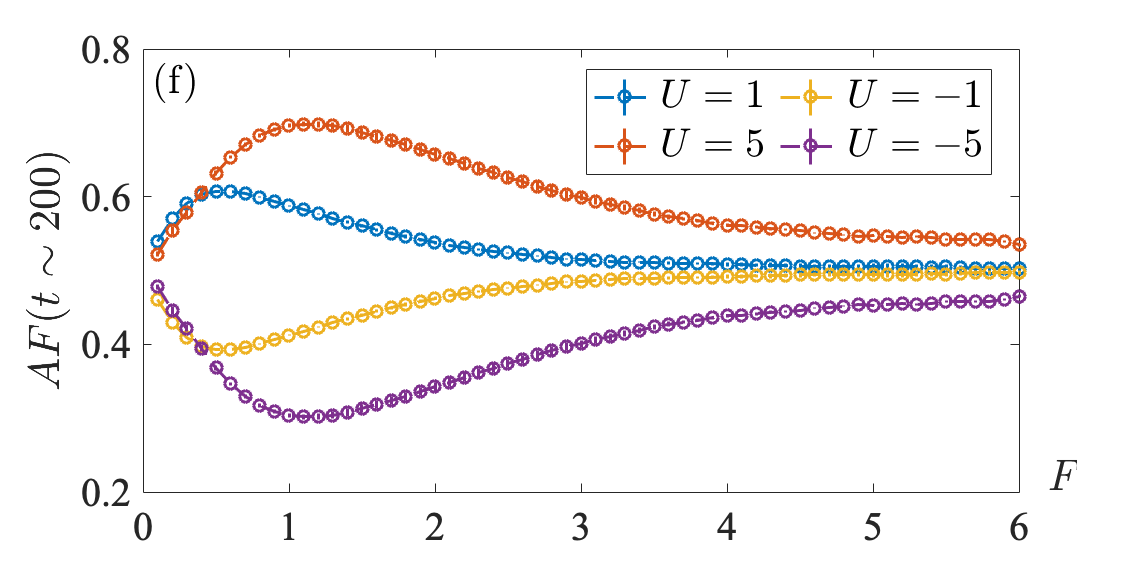}\vspace{-0.2cm}
\includegraphics[width=1\linewidth]{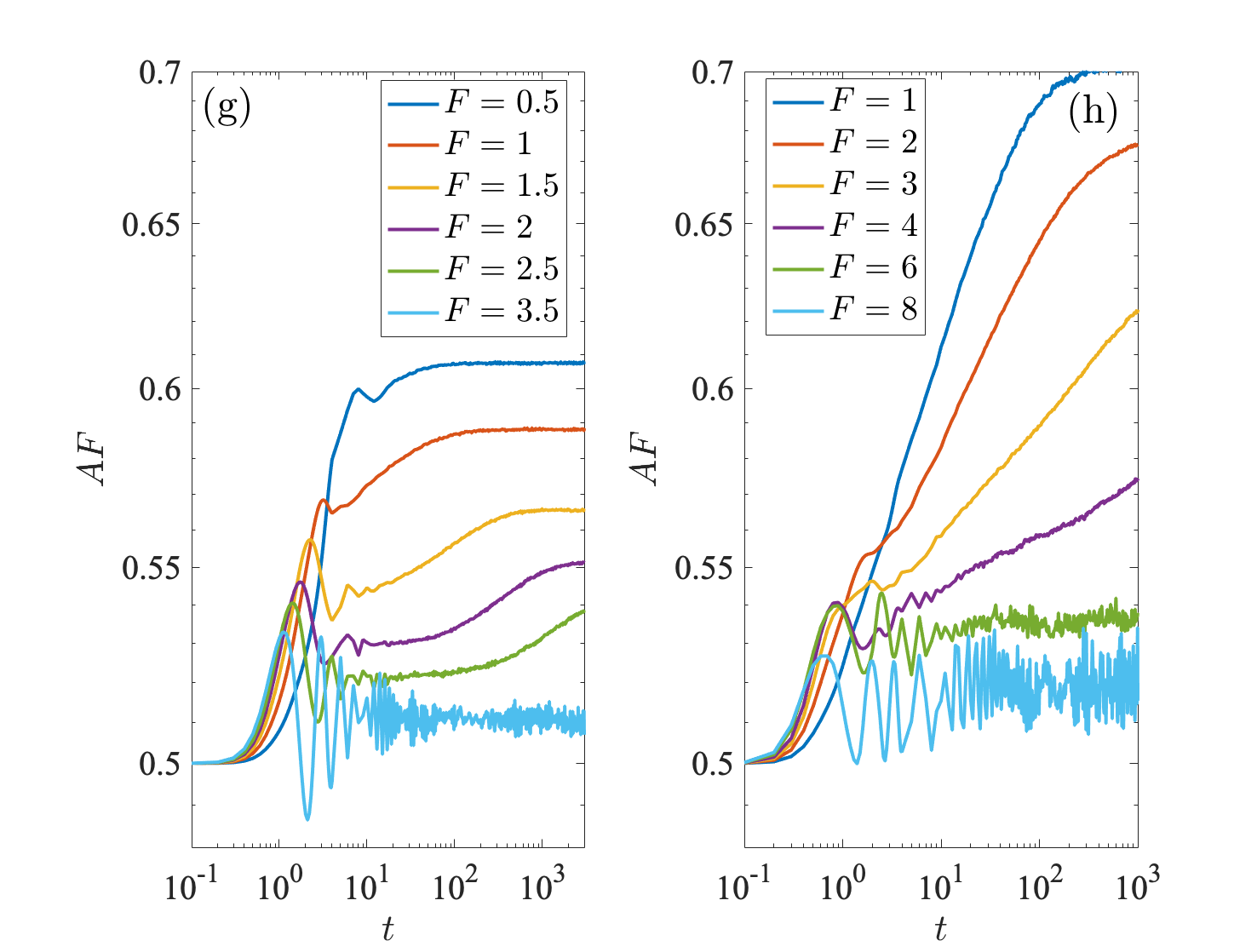}\vspace{-0.3cm}
\caption{{(a) Schematic plots for time evolution. (b-e) The final occupations under different strength of the tilt{, $F$}. 
{Panels correspond to different interaction strengths: (b) -- $U=1$; (c) -- $U=5$; (c) -- $U=-1$; (d) -- $U=-5$ Negative interaction case results could simply be obtained by reversing the lattice. (f) Accumulation Factor depending on strength $F$, maxima corresponds to enhanced transport. We visualize transport behavior by looking at long time dynamics of accumulation factor: (g) $U = 1$, (h) $U = 5$.}
    \label{linear_dynamics} \vspace{-0.15cm}
}}
\end{figure}

While spectral properties characterise the system in quite a complete way, in view of possible experiments we address the time dynamics in the system. For small number of sites in bosonic system it was shown \cite{Rispoli19} that a complete characterization of occupation of sites and their correlations is accessible. With these impressive results in mind we
consider the time dynamics for an unit average filling per site. Using 
Chebyshev propagation scheme \cite{Tal-Ezer84,Cheby91,Fehske09} we can reach $L=12$ sites with $M=12$ bosons, the system not easily accessible for direct diagonalization.
As the initial stare we consider an uniformly filled Fock state with one particle at each site:  $| 1,1,..., 1 \rangle$ state, then observe its time evolution in a tilted lattice.  We collect the final occupation pattern around $t \sim 200$ for $F \in (0,6]$ as shown in Fig.~\ref{linear_dynamics}. 
For $F=0$ the initially symmetric in space state remains symmetric with nearly uniform occupations except at the edges (we consider open boundary conditions), as we know from earlier studies \cite{Yao20} the state becomes significantly entangled during time evolution. The spacial symmetry is broken for non-zero $F$.  Small  $F$ values lead to transport of bosons which accumulate at one side due to the tilt of the optical lattice. This behaviour is reversed for larger $F$ values, when the localization sets {in and}  the almost  uniform occupation of sites is observed even after a long time -- see  Fig.~\ref{linear_dynamics}(b) for $U=1$.  For stronger interactions $U=5$, {Fig.~\ref{linear_dynamics}(c),} one could naively expect that bosons repel stronger
and the uniform site distribution is created for smaller $F$. This is not the case, we see that even for the strongest considered value $F=6$ a slight asymmetry remains close to the edges of the system - it correlates well with  the gap ratio behavior in Fig.~\ref{rbarheis} where, for strong interaction case the border of MBSL is shifted to $F>5$.

The attractive interaction case -- Fig.~\ref{linear_dynamics}(d-e) is pretty interesting. Contrary to the intuition  the bosons, outside of the localization regime,  move against the potential and accumulate at high potential energy end. The picture is just a mirror image of the $U>0$ behavior due to the symmetries of the Hamiltonian discussed above.

\par
To quantify the degree of net transport induced by the tilt we define the Accumulation Factor (AF) as: 
\begin{equation} 
AF = \frac{\sum_{i \in [1,L/2]} n_i}{\sum_{i \in [1,L]} n_i}
\label{AF}  
\end{equation}
so that $AF \in [0,1]$ measures a fraction of particles occupying  the left hand half chain with $AF=0.5$ corresponding to the same mean occupation of left and right half chain. 
This is the case for $F=0$ as well as when a strong MBSL sets in.  $AF$ can be measured at arbitrary $t$, we present its $F$ dependence at the exemplary final time $t=200$ for all four different $U$ cases considered in  Fig.~\ref{linear_dynamics}(f). For positive $U$ both curves reveal an initial growth and then the decay when MBSL sets in. For $U=1$ around $F\approx4$ $AF$ comes back entirely to 0.5 value - this correlates again very well the the gap ratio statistics. For $U=5$ apparently MBSL is not complete even at $F=6$.
\par

{To analize the transition to MBSL by time dynamics, we plot $AF(t)$ in Fig.~\ref{linear_dynamics} for $U=1$ (g) and $U=5$ (h), respectively. For
$M=12$ bosons on $L=12$ sites  we follow the time dependence  of $AF(t)$  up to $3000(1000){\hbar}/{J}$, respectively in the logarithmic scale. The $F$ values taken are below full MBSL case and one may clearly identify  three regimes: (i) a fast initial redistribution of particles on the time scale of few tunneling times; (ii) almost linear growth (on the logarithmic scale) corresponding to slow subdiffusive-like growth (iii) saturation when the quasi-stationary distribution is reached (for small system sizes considered). This behavior of
$AF(t)$ resembles to a large extend the time dynamics of the transport distance analysed in the transition to MBL in \cite{Rispoli19,Yao20}. The latter quantity requires two-point
correlation function evaluation while $AF(t)$ relies on occupations only.}

{Observe that the data for larger $f$ values, corresponding to localization with low $AF$ appear more noisy in Fig.~\ref{linear_dynamics}. This is due to the fact that apparently few eigenstates contribute significantly to the evolution of the initial wavepacket resulting in the quasiperiodic oscillation of observables. Such oscillations look quite irregular on a logarithmic scale. }

 \begin{figure}
\includegraphics[width=0.85\linewidth]{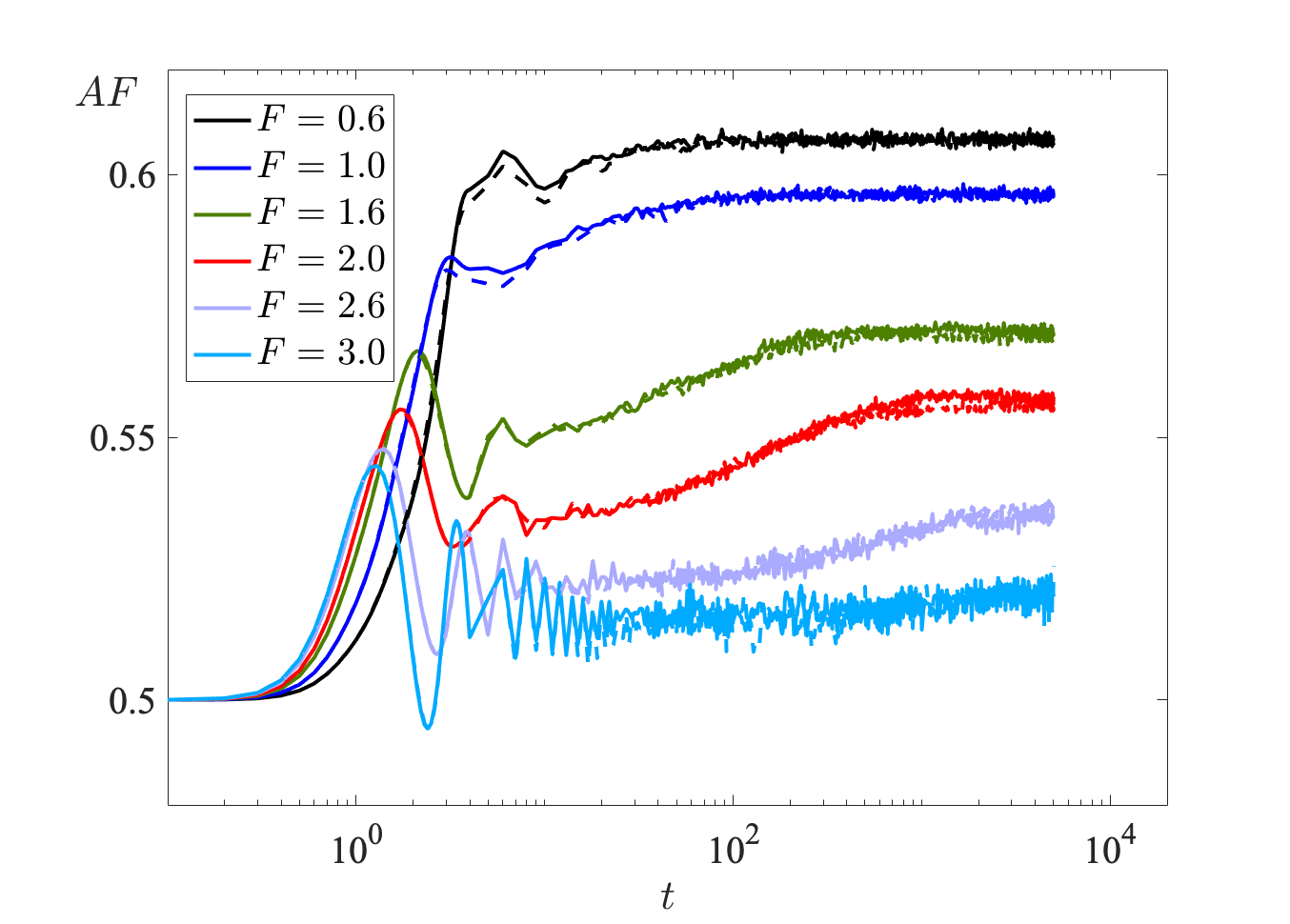}
\caption{Comparison of time dynamics for $M=10$ particles on $L=10$ sites starting from unitial uniform Fock state for different $F$ values as indicated in the figure. Dashed lines 
correspond to the evolution in the presence of additional harmonic trap, adding to the chemical potential the term $h_i=\frac{A}{2}i^2$ with $A=0.033$ so the tilt of the lattice remains approximately the same. The effect of the additional harmonic term on time dynamics is negligible on the time scale considered.
    \label{fracton}
    }
\end{figure}

{It has been noted \cite{Schultz19,Taylor19} that the system in tilted lattice belongs to a class of systems with global constrains, not only the charge (the particle number) but also the dipole moment is conserved. For such systems fracton excitations are claimed to be responsible for eventual thermalization of the system at very long time due to very slow dynamic of hydrodynamical origin \cite{Feldmeier20,Gromov20}. To overcome these effects and observe truly localized systems additional small terms to the Hamiltonian are added 
as a small disorder \cite{vanNieuwenburg19} or a small additional harmonic potential at sites adding to $\mu_i$ in \eqref{chempot} the additional term $\sim i^2$. While such an approach is necessary for level spacing analysis (due to quasi-degeneracies in the spectrum for pure Stark problem \cite{Schultz19,Taylor19}) the fracton dynamics does not occur on the experimental time scale as shown by comparison of the dynamics with and without the additional harmonic term -- Fig.~\ref{fracton}.

\section{Bosons in a tight harmonic trap}
The harmonic trap in some form typically accompanies the optical lattice potential. Typically in experiments the curvature of the potential is quite tiny, promising that systems to be studied are locally uniform. Even then it may lead to coexistence of different phases as exemplified by the famous cake shape for the ground state occupation of bosons within Bose-Hubbard model where a harmonic potential modifies local chemical potential creating regions of Mott insulator and superfluid phases (for a review see \cite{Bloch08}). By shaping light with digital micromirror devices  \cite{Gauthier16, Mazurenko17} or spatial light modulators  \cite{Gaunt_Thesis} one can remove the undesired remaining trapping potentials  or add an arbitrarily designed envelope to the system studied.
 Modifying the curvature  one
may improve adiabatic loading of cold atoms into the ground state \cite{Rey06,Zakrzewski09} and have access to the compressibility of the sample \cite{Delande09}.

Recently {the dynamics of excited states}  in the presence of a harmonic trap on top of the optical lattice potential has also been  studied. It was demonstrated that extended and localized phases may coexist for spinless as well as spinful fermions  \cite{Chanda20b}. Hereby, we test the  same idea for bosons assuming the chemical potential in the form.
\begin{equation} 
\mu_i = \frac{A}{2}(i-i_o)^2=\frac{A}{2}(i-\frac{L+1}{2})^2,
\label{muhar}  
\end{equation}
where $i_0$ is the center of the trap.
\begin{figure}
\includegraphics[width=1\linewidth]{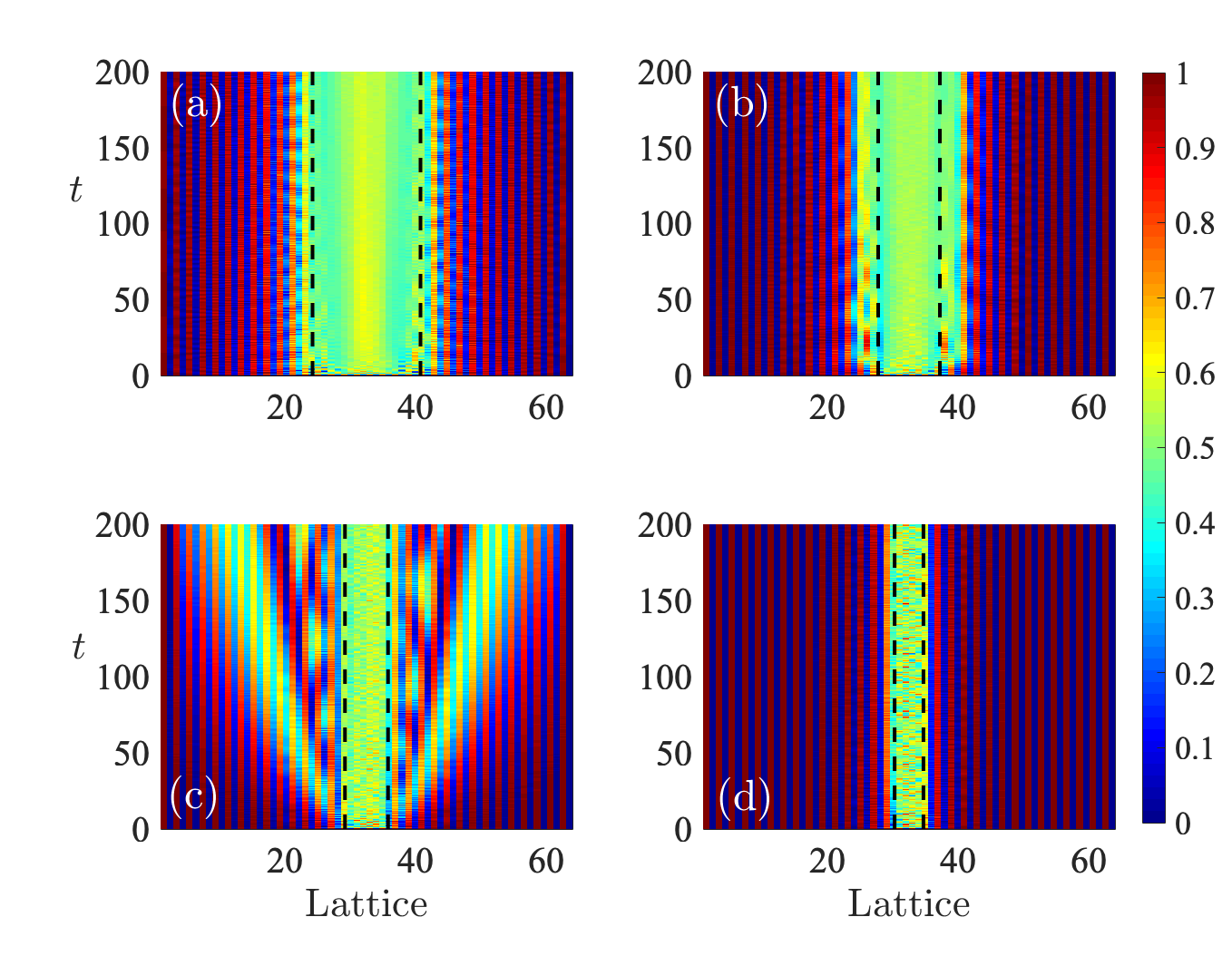}
\caption{{Time evolution of site occupations in optical attice suplemented by a harmonic trap with different harmonic curvatures $A$: (a)$A = 0.4$. (b)$A = 0.7$. (c)$A = 1.0$. (d)$A = 1.5$. All plots are for interacting.bosons with $U = 1$. Dashed lines are indicating the  coexistence boundaries  obtained assuming a local field argument (see text)  $i_c = i_0\pm F_c/A$ with the  critical field $F_c = 3.3$. 
    \label{harocc}
}}
\end{figure}
\par
We visualize the coexistence phenomenon induced by harmonic trap by considering the time evolution of a staggered initial state $|\psi\rangle = |0,1,0,1,...\rangle$ in a chain with 64 sites. {To treat such a relatively large system at half filling we use time dependent variational principle (TDVP) algorithm(for a review of numerical tools enabling study of time dynamics 
for large system sizes see \cite{TDVP_review}). In the simulation we assume maximal number of bosons per site, $n_{\rm max} = 6$ using typically auxilliary space dimension $\chi = 256$ and $\chi = 384$. The simulation is performed with time step $0.05{\hbar}/{J}$ and cut-off $10^{-9}$. Tests of convergence show that the results are reliable for times  considered in later discussion.} At sufficient long time, the middle of chain appears to be thermalizing and the initial staggered occupations spread over the central region - compare Fig.~\ref{harocc}. However, in both  outer regions the system preserves the memory  of its original configuration showing the lack of thermalization - localization occurs. By increasing the curvature $A$, the boundary separating the apparently coexisting localized and thermalized  regions moves towards the center.
 
As observed by us for fermions \cite{Chanda20b} an understanding of this behavior may be obtained invoking the notion of the local field $F(x) = \frac{d\mu(x)}{x} \approx Ax$. If the local field in a given region is sufficiently strong so it  would lead to localization in a tilted lattice, one may expect localization in this region (of sufficiently large curvature).  For fermions, there is a strict correspondence between the critical field $F_c$ leading to localization and 
the local field $F=A(i-i_0)$ leading to a separation of delocalized center of the trap from the localized sides for which $|i-i_0|>A/F_c$. The same approach, using the critical field
$F_c\approx 3.3$ obtained previously (compare Fig.~\ref{Rescale}), yields  dashed lines estimates that are close to the boundary of two phases, supporting our local field hypothesis.
The central delocalized region exceeds a little the local field borders for $A<1$.
\par
For $U = A = 1$ (Fig.~\ref{harocc}(c)) there is an unexpected pattern in time evolution: an excitation emerges from the center, penetrates the localized region moving outwards with a small spread. This resonance-like effect may be understood (we are grateful to Piotr Sierant for contributing his insight to this point) considering the family of states $|\psi_j\rangle$ that, for $U=A$ 
are degenerate with the initial state $ |\psi_0\rangle = |1,0,1,0,1,0,1,0,1,0...\rangle$:
\begin{equation}
\begin{aligned}
& |\psi_1\rangle = |0,2,0,0,1,0,1,0,1,0...\rangle \\
& |\psi_2\rangle = |1,0,0,2,0,0,1,0,1,0...\rangle \\
& |\psi_3\rangle = |1,0,1,0,0,2,0,0,1,0...\rangle \\
& \vdots
\end{aligned}
\end{equation}
as the  energy difference between state $|\psi_j\rangle$ and state $|\psi_0\rangle$ is $\Delta E_{0,j} = \mu_{2j-1} + \mu_{2j+1} - 2\mu_{2j} - U = A-U$. 
Within
this degenerate subspace $|\psi_0\rangle$ is coupled to $|\psi_j\rangle$ by a two-fold action of the hopping term, i.e.boson on 
site $2j-1$ hops onto site $2j$ and boson from $2j+1$ hops onto $2j$. The two processes sum up to a
 second order process
occuring at a position-depending rate 
\begin{equation}
\begin{aligned}
r_j = &\frac{J^2}{(\mu_{2j}-\mu_{2j-1})} + \frac{J^2}{(\mu_{2j}-\mu_{2j+1})} \\
= &\frac{J^2}{(2j-i_0-1/2)(2j-i_0+1/2)A}
\end{aligned}
\label{rate} 
\end{equation}
 where, recall, $i_0$ is the trap center.
To simplify the argument let us count the index from the center of the trap, effectively shifting $i_0=1/2$. We get then the decreasing 
series of effective hopping rates $r_j=J^2/2j(2j+1)A$  forming a series of time scales $t_j\sim 1/r_j$ for entangling
 $|\psi_0\rangle$ with $|\psi_j\rangle$. 
 Since $j$ is nothing as a discretized distance from the center of the trap, we have a parabolic dependence linking time $t$ with distance $t\sim 4j^2A/J^2$.
 Such a parabola is indeed observed in Fig.~\ref{harocc}(c).
 The resonance occurs for arbitrary $U$ once $A=U$ condition is satisfied as shown in Fig.~\ref{scale_emission}.
 
\begin{figure}
\includegraphics[width=1\linewidth]{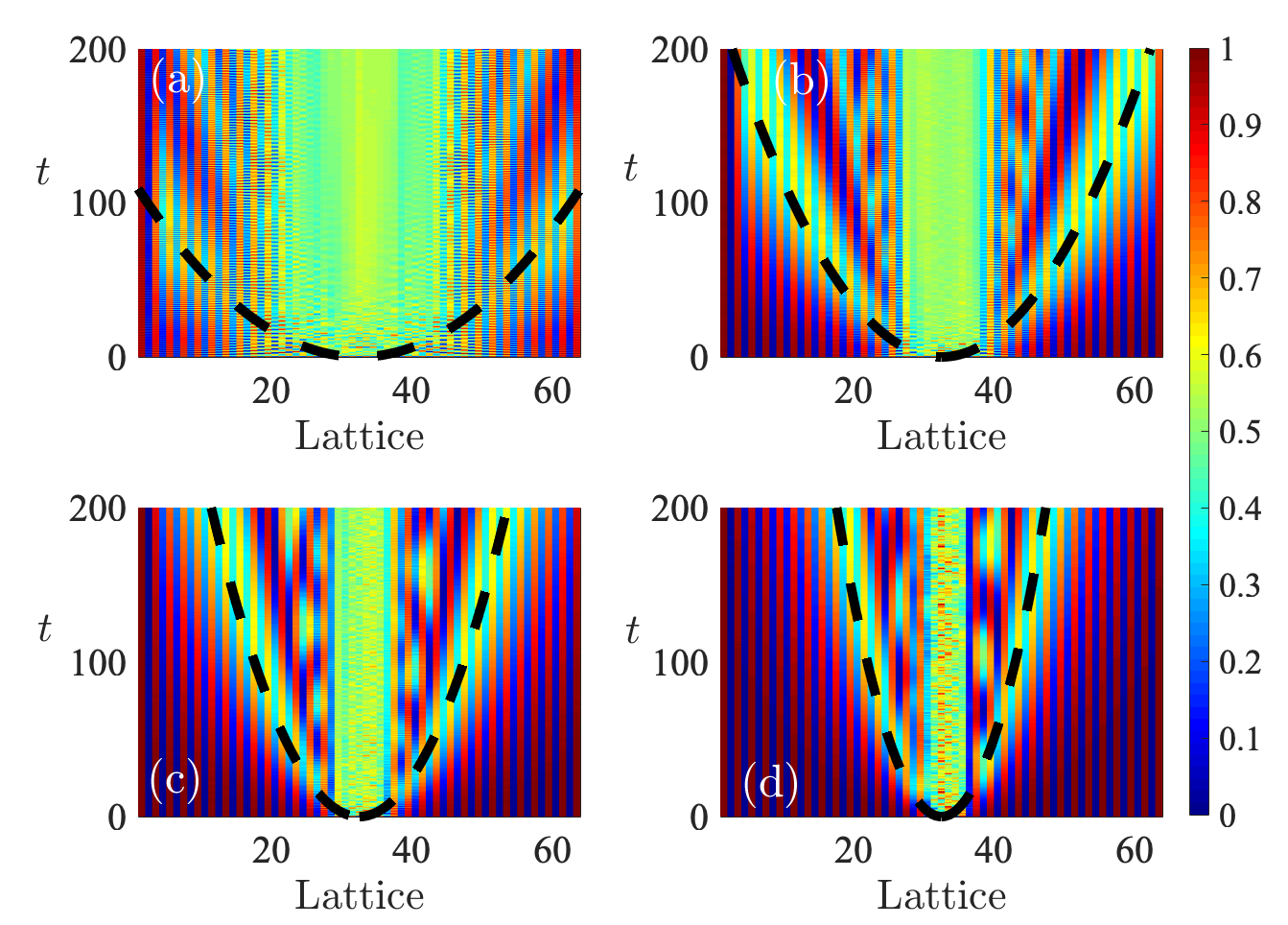}
\caption{{The resonance under different interaction strengths: (a)$U = A = 0.2$. (b)$U = A = 0.5$. (c)$U = A = 1.0$. (d)$U = A = 2.0$. The speed of emission changes depending on interaction $U$. Black dashed curves depict parabolic fitting to emission patterns $t = C(s-i_0)^{2}$ with C proportional to A as discussed in the text.
    \label{scale_emission}
}}
\end{figure}

Observe that the prominent ``parabola'' excitation, described above,  is followed for $A=U$, by additional emissions creating small local grains of roughly half-integer populations.  Those grains seem to lay on another parabola's with a slower spread. We believe that they are due to higher order processes within the discussed degenerate manifold. 
\par
The entanglement entropy growth follows the occupations pattern, again in a close similarity with the fermionic case \cite{Chanda20b}. In the central delocalizing region the entropy (as measured on different bonds) grows rapidly and saturates, while in the localized outer regions it exhibits logarithmic growth as in SMBL case - compare Fig.~\ref{entroemission}. A closer inspection of Fig.~\ref{harocc} and the upper row of Fig.~\ref{entroemission} shows interesting feature. While occupations redistribute themselves 
very fast and the occupations practically equalize in the central region on the time scale of few tunneling times, the entanglement entropy shows a different behavior. It grows fast for the central bond but the growth is much slower at bonds say 28 and 24 which are within the thermalizing region. In effect, the region of large entropy spreads slowly in time, staying well within the borders given by the local field estimate. Such a slow down of the  growth of the entanglement entropy inside the thermalizing center was observed already for
fermions \cite{Chanda20b}. It was attributed to the fact that due to small local Hilbert space for fermions there are limitations on the difference between entanglement entropies on nearby bonds. Thus a slow growth on the localized side affects also the growth in the central region. Interestingly we observe the same behavior for bosons for which, in principle, the dimension of the local Hilbert space is unlimited (it is limited in our calculations to $n _{\rm max} = 6$ but we have checked that the increase of $n_{\rm max}$ does not affect the time dynamics of entropy).
 
For the resonance case, the emission effects are visible also in the entropy growth - we observe an oscillatory dynamics imposed on the growth. $S(t)$ peaks when the excitation propagates across the bond to be considered, and the subsequent emissions lead, similarly, to additional oscillations.
\begin{figure}
\includegraphics[width=1\linewidth]{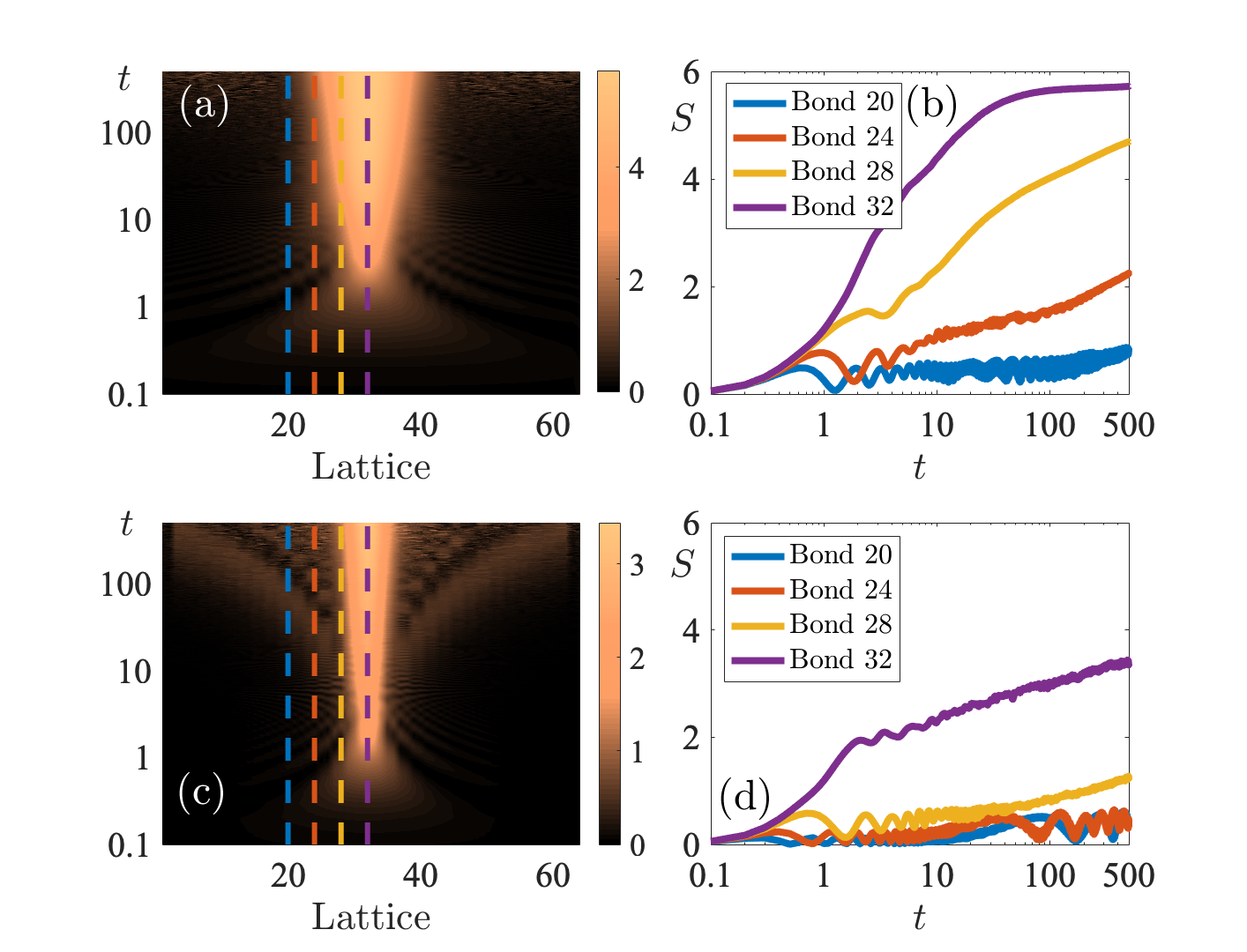}
\caption{{The time dynamics of the entanglement entropy on different bonds for $U=1$ and $A = 0.4$ (upper panel) and the resonant case $A = 1$ (lower panel). The auxiliary space dimensions are set to be $\chi = 384,256$ respectively, ensuring the convergence of the results up to $t = 500$. The resonance occuring for $A = 1.0$ results in oscillations in $S(t)$. 
   \label{entroemission}
}}
\end{figure}
\par

To end this section we present the evidence for the convergence of our simulations by considering the entanglement entropy at different bonds for different auxiliary space dimension $\chi$. We investigate $\chi = 64,128,256,384$ and extract $S(t)$ from the same bonds as in Fig.~\ref{entroemission}. The weaker the harmonic trap is, the larger $\chi$  is required for a given accuracy (the thermalizing central region is bigger). Therefore, we show $A = 0.4$ and $A=0.7$ cases -- see Fig.~\ref{Dcut}. For $A = 0.7$, $\chi = 384$ provides a satisfactory convergence for all bonds up to $t=500$, while for $A = 0.4$ even such a large $\chi$ value is insufficient for the central $b=32$ bond. Observe that the convergence is restored quite fast when moving away from the very center of the trap, even well within the thermalizing center. Since we do not analyse in detail properties of the system  in the very center of the trap  a simulation with $\chi = 256$ is already a good choice for $A=0.7$ or larger while for $A = 0.4$ $\chi = 384$ is definitely required. As it is clear from lower panels in Fig.~\ref{entroemission} also in the resonant case, despite travelling excitations, the entropy growth is limited and may be reliably simulated with $\chi=256$.
\begin{figure}
\includegraphics[width=1\linewidth]{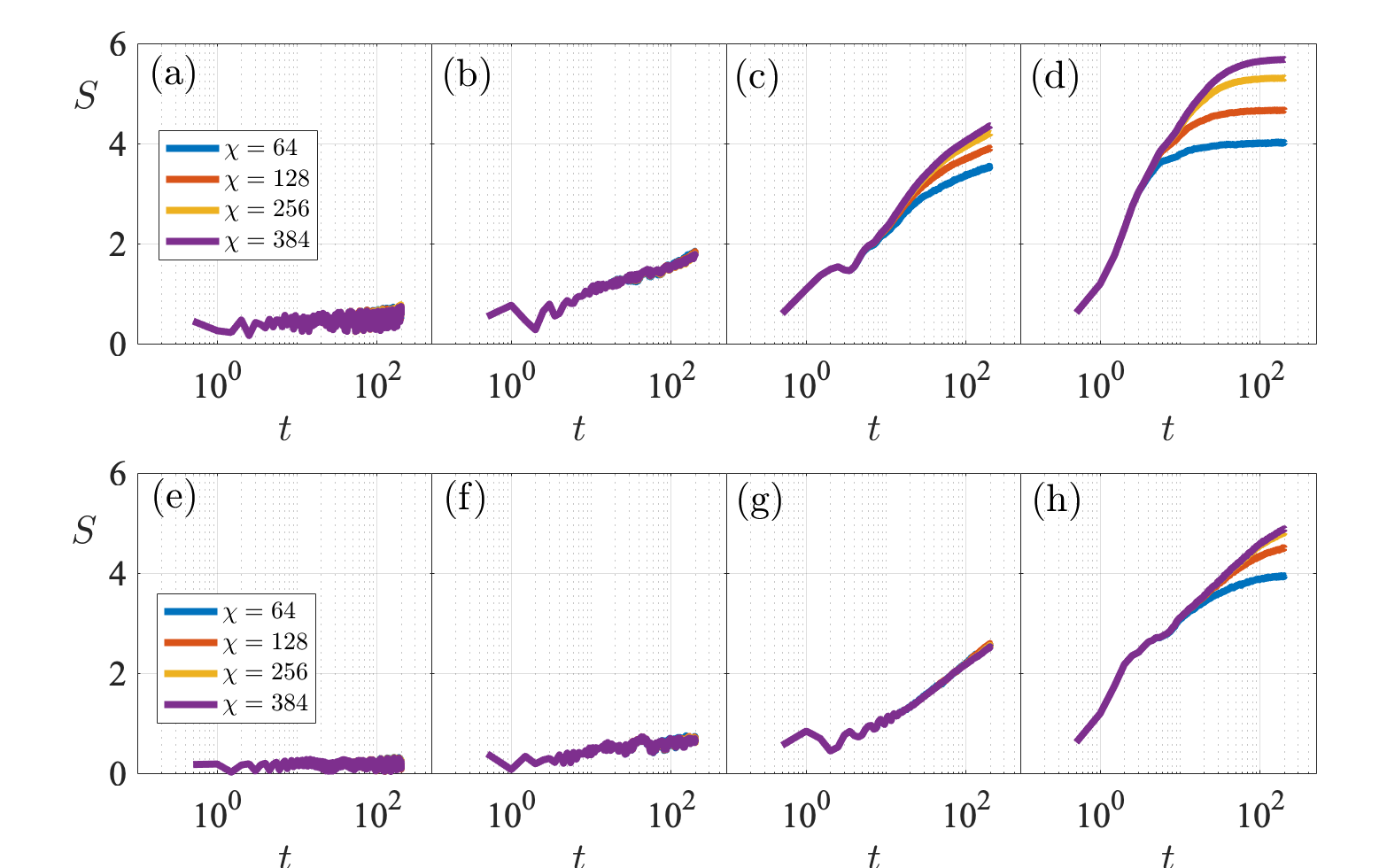}
\caption{{Entanglement entropy time dynamics on different bonds for different auxiliary space dimensions as indicated in the figure. Upper panel corresponds to $A = 0.4$ while the lower one is for $A = 0.7$. Except at the very middle of the chain for $A\le 0.4$, the entropy is reasonably converged  for $\chi = 256$. 
   \label{Dcut}
}}
\end{figure}

\section{Inverted trap and its confinement}
We have observed that the harmonic potential on top of the optical lattice could induce coexistence of localized and  thermal phases strongly suppressing the transport between these domains. The effect is due to local effective electric fields that, if exceeding the threshold value, lead to localization. The effect does not depend on the {\it sign} of the curvature, as what really matters is the local field. In effect, an inverse harmonic
trap should also be able to prevent atoms from expansion and loss. The effect is entirely of different origin from the fact that, a long-lived {\it attractively interacting} bosons may be confined in inverse trap as demonstrated by \cite{Braun13} as a 
consequence of negative temperature. In our case the confinement is due to localization induced suppresion of transport and is independent (or weakly dependent) on the sign of the interaction.
\par
Let us demonstrate the effect in a chain of size $L = 64$ and consider as the initial state  a pure state with middle 14 sites occupied by one particle each with the rest of the chain being empty. This configuration is, with no doubt, unstable and all particles should expand with repulsive interaction $U > 0$ while shrinking for sufficiently large $U < 0$. The simulation could be considered as a simplified version for an expansion of initially well-confined atomic gas. For attractive interactions $U = -1$, the occupations evolved with time under reversed trap $A = -0.4,-1$ are depicted  in the top row of Fig.~\ref{Antitrap}. The atoms accumulate in the center - the simulation reflects simply earlier  experimental results  \cite{Braun13}, indicating a long-lived trapped mode interpretted  as the negative temperature effect. However, for $U = 1$, while intuitively the gas should expand across the lattice, the expansion is stopped when the local field $F_i = A|i-i_c|$ reaches critical value. Atoms are ``forbidden'' to enter the localized regime - the wavepacket has apparently vanishingly small overlap on 
eigenstates strongly localized in the outer region.  As revealed by inspection of Fig.~\ref{Antitrap}, the difference between the interactions being attractive or repulsive shows as the distribution of particles in the central region: for attractive case they tend to occupy the very center forming a  is single peak  but for the repulsive case there is an excess populations close to the boundaries of the central region. 
\begin{figure}
\includegraphics[width=0.95\linewidth]{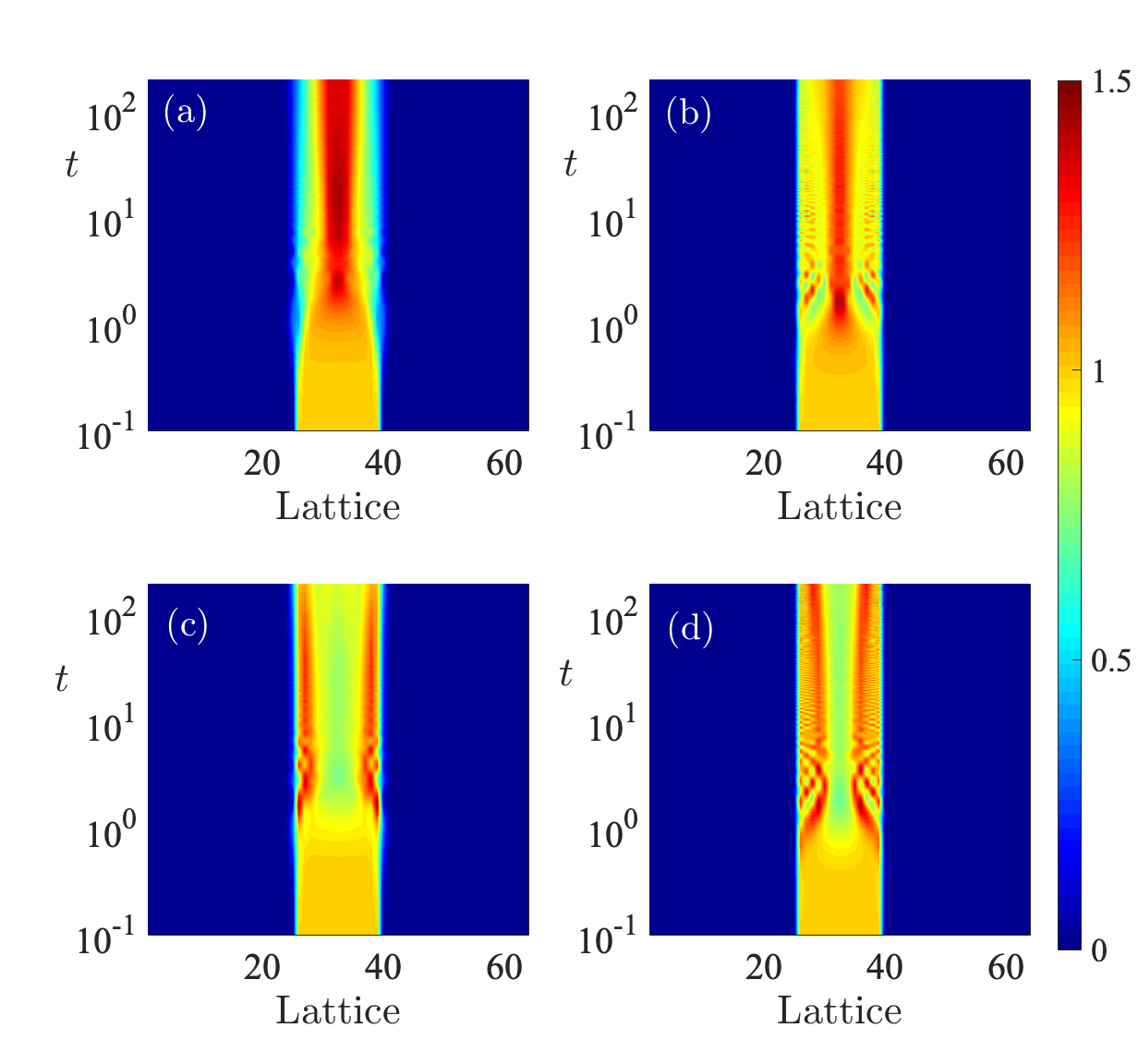}
\caption{{Time dynamics of a small bosonic cloud initially occupying with unit filling $M=14$ central sites for different combinations of interactions and curvatures values (a) $U = -1, A = -0.4$;  (b) $U = -1, A = -1$;  (c) $U = 1, A = -0.4$;  (d) $U = 1, A = -1$. No particle moves outside into the ``localized'' regions.
    \label{Antitrap} 
}}
\end{figure}

\section{Conclusions}
We have shown that interacting bosons in optical lattice may be many-body localized  in the presence of a local force $F$ in similarity with spinless and spinful fermions. The resulting Stark many-body localization is similar to disorder induced MBL - in particular eigenstates in the localized regime show multifractal properties. The Stark MBL has an impact on the behavior of interacting particles in an arbitrary potential - we demonstrate in detail the system dynamics in the presence of the harmonic trap. Then the coexistence of apparently thermalizing region with outer regions exhibiting strong localization has been demonstrated, in analogy to the similar behavior observed for fermions 
\cite{Chanda20b}. The border separating localized and thermal parts is, to a good precision, given by the critical value of the  static field (force) which leads to Stark many body localized system in the tilted lattice. Such a local field value is given by the spacial derivative of the potential (not necessarily harmonic) so the effect should not be limited to harmonic potential but it is rather a generic feature of slowly varying potentials. As an example we show  that even in the inverted harmonic trap which is supposed to loss atoms rapidly -- surprisingly no losses appear and the atomic cloud is well confined  as a consequence of suppression of transport into the many-body Stark localized neighboring regions. More complicated in shape potentials may separate the space into several regions with transport practically prohibited 
between them.

Our numerical results are either related to small systems amenable to exact treatment (via diagonalization or Chebyshev propagation) or to typical experimental times of hundreds of tunneling times (for TDVP simulations). This does not preclude that, for example, the coexistence of localized and thermal regions very slowly fades away in the large systems/long times
limit. Additional studies are needed to resolve those issues. Such studies are, however, at the border of current numerical capabilities.

 \begin{acknowledgments} 
 We are grateful to Titas Chanda and Piotr Sierant for discussions on different aspects of this work and remarks on the manuscript.
The numerical computations have been possible thanks to  High-Performance Computing Platform of Peking University. Support of PL-Grid Infrastructure is also acknowledged. The TDVP simulations have been performed using ITensor library (\url{https://itensor.org}). 
This research has been supported by National Science Centre (Poland) under project  2019/35/B/ST2/00034  (J.Z.).
 \end{acknowledgments}

%

\end{document}